\documentclass[conference]{IEEEtran}
\IEEEoverridecommandlockouts
\usepackage{cite}
\usepackage{amsmath,amssymb,amsfonts}
\usepackage{algorithmic}
\usepackage{graphicx}
\usepackage{subfigure}
\usepackage{textcomp}
\usepackage{multirow}
\usepackage[table]{{xcolor}}
\usepackage{xcolor}
\def\BibTeX{{\rm B\kern-.05em{\sc i\kern-.025em b}\kern-.08em
    T\kern-.1667em\lower.7ex\hbox{E}\kern-.125emX}}
\begin{document}

\title{Exploring the Relationship Between Personality Traits and User Feedback
}

\author{\IEEEauthorblockN{Volodymyr Biryuk and Walid Maalej}
\IEEEauthorblockA{\textit{Department of Informatics} \\
Universität Hamburg \\
Hamburg, Germany \\
firstname.lastname@uni-hamburg.de}
}

\maketitle

\begin{abstract}
Previous research has studied the impact of developer personality in different software engineering scenarios, such as team dynamics and programming education.
However, little is known about how user personality affect software engineering, particularly user-developer collaboration. 
Along this line, we present a preliminary study about the effect of personality traits on user feedback.  
56 university students provided feedback on different software features of an e-learning tool used in the course. They also filled out a questionnaire for the Five Factor Model (FFM) personality test. 
We observed some isolated effects of neuroticism  on user feedback: most notably a significant correlation between neuroticism and feedback elaborateness; and between neuroticism and the rating of certain features.
The results suggest that sensitivity to frustration and lower stress tolerance may negatively impact the feedback of users.
This and possibly other personality characteristics should be considered when leveraging feedback analytics for software  requirements engineering. 
\end{abstract}

\begin{IEEEkeywords}
user feedback, personality types, personality traits, requirements engineering
\end{IEEEkeywords}

\section{Introduction}

Personality types, which are combinations of personality traits, have been studied for decades---not only in psychology but also in software engineering.
Usually, a variety of personality tests such as Myers-Briggs Type Indicator (MBTI)~\cite{briggs1976myers}, Big Five/Five Factor Model (BF/FFM)~\cite{costa1992revised}, Keirsey Temperament Sorter (KTS)~\cite{keirsey1998please}, Sixteen Personality Factor Questionnaire (16 PF)~\cite{cattell1970handbook}, Adjective Check List (ACL)~\cite{gough1983adjective}, or Eysenck Personality Inventory (EPI) \cite{zeigler-hillEncyclopediaPersonalityIndividual2020} are used to group participants into different personality types~\cite{cruzFortyYearsResearch2015}.

In software engineering, research has focused so far on studying the personality types of developers and their impact  on development performance, team dynamics, leadership performance, and task allocation\cite{cruzFortyYearsResearch2015}.
The evidence suggests, e.g., that certain personality types prefer certain software engineering tasks and that the correct distribution of tasks has an impact on project success~\cite{cruzFortyYearsResearch2015}.
However, the success of software engineering projects is also largely dependent on the contribution of other stakeholders, most notability of users during requirements engineering and maintenance activities. 
Yet, little is known on whether and how user personality impact these activities.  

In recent years,  the systematic collection and analysis of user feedback to inform requirements engineering and software evolution tasks has become a common practice \cite{wangSystematicMappingStudy2019a}.
Especially, since the emergence of social media and  app stores, a significant amount of feedback on software can be collected and used by development teams, who also increasingly engaging in conversations with users \cite{Martens:RE:2019}. 
Feedback can reveal the users' perspective on the software and its features that is not obvious to the developers, for example unexpected usage scenarios, unreported defects, or ideas for new features \cite{paganoUserFeedbackAppstore2013}.
This work explores whether personality types can effect the  feedback behavior of software users. 
If yes, personality traits should be taken into account to ``calibrate'' and interpret feedback collected online, in workshops, and interview sessions. 

Our intuition is that the personality type might have an impact on how individuals perceive software, and consequentially get engaged, and frame their feedback.
For example, users with a certain, highly developed personality trait may get frustrated with certain features quicker than others or focus more on the negative aspects.
Other users might feel obliged to explain their position, or are able to spot aspects of the software that others do not.
This can be reflected in different properties of feedback such as software rating, wording, scope,  mentioned features, or motivation and expectation.
If such relationships exist, they should be taken into consideration in the collection, prioritization, and handling of feedback.


We report on a preliminary study in the context of an introductory programming course, where we analyzed user feedback given by university students to the  online teaching platform Moodle together with the personality traits of the students. 
Our research questions are as follows:

\begin{itemize}
    \item \textbf{RQ1}: Do feature ratings correlate with personality trait scores?
    \item \textbf{RQ2}: Do personality trait characteristics correlate with the elaborateness of user feedback?
\end{itemize}

We introduce the background for personality studies and close related work in Section \ref{sec:background}. 
Then, we present the design of our study in Section \ref{sec:method}, the results in Section \ref{sec:results}, and threats to validity in Section \ref{sec:threats}. 
Finally, we briefly discuss the findings in Section \ref{sec:discussion} and conclude the paper in Section \ref{sec:conclusion}.

\section{Background}\label{sec:background}

\subsection{Personality Traits and Types}
Personality traits are relatively enduring patterns in behavior, thoughts, and feelings, that manifest the tendency to respond in certain ways under certain circumstances \cite{robertsBackFuturePersonality2009}.
The constellations of those unique personality trait features are called personality types and are used to define discrete groups of individuals \cite{costaReplicabilityUtilityThree2002}.
In personality psychology, they are often used to better understand the diversity of responses from people in similar circumstances \cite{borghansIdentificationProblemsPersonality2011}.

Three personality tests dominate the scientific studies so far: namely Myers-Briggs Type Indicator (MBTI) \cite{briggs1976myers}, Five Factor Model (FFM), and Keirsey Temperament Sorter (KTS) \cite{cruzFortyYearsResearch2015}. 
The FFM model is the most suitable for our research goal, as it produces more comprehensive information on all scales~\cite{furnhamBigFiveBig1996, mccraeContemplatedRevisionNEO2004}.
Furthermore, MBTI lacks the representation of neuroticism~\cite{furnhamBigFiveBig1996}, which is an important dimension when considering  frustration during software usage and the subsequent feedback behavior.
The FFM includes the following five personality traits: 

\begin{itemize}
\item \textbf{Agreeableness (A)}
Individuals scoring high on this dimension tend to be compliant, cooperative, altruistic, and helpful to others \cite{roccasBigFivePersonality2002, rothmannBigFivePersonality2003}. 
Individuals who score low on this dimension tend to be irritable, skeptical towards others, competitive, and egocentric \cite{roccasBigFivePersonality2002, rothmannBigFivePersonality2003}.
    
\item \textbf{Agreeableness (A)} Individuals scoring high on this dimension tend to be compliant, cooperative, altruistic, and helpful to others~\cite{roccasBigFivePersonality2002, rothmannBigFivePersonality2003}. 
Individuals who score low on this dimension tend to be irritable, skeptical towards others, competitive, and egocentric~\cite{roccasBigFivePersonality2002, rothmannBigFivePersonality2003}.

\item \textbf{Conscientiousness (C)}
Individuals scoring high on this dimension tend to be careful, thorough, responsible, organized, and forward planning \cite{roccasBigFivePersonality2002, rothmannBigFivePersonality2003}.
Those who score low on this dimension tend to be irresponsible, disorganized, and unscrupulous~\cite{roccasBigFivePersonality2002, rothmannBigFivePersonality2003}.
High levels of conscientiousness manifest themselves in an achievement-oriented and dependable character.
Low levels of conscientiousness, however, do not imply a lack of work ethics, but rather a lower ability to apply them~\cite{rothmannBigFivePersonality2003}.

\item \textbf{Extraversion (E)}
Individuals who score high on this dimension tend to be sociable, talkative, assertive, and sociable~\cite{roccasBigFivePersonality2002, rothmannBigFivePersonality2003}.
Introverted individuals (those who score low on this dimension) are often retiring, reserved, cautious, and independent~\cite{roccasBigFivePersonality2002, rothmannBigFivePersonality2003}.

\item \textbf{Neuroticism (N)}
Individuals who score high on Neuroticism have a tendency to experience sadness, embarrassment, anxiety, depression, and anger. They are prone to insecurity and irrational ideas, are less able to control impulses, and cope poorly with stress~\cite{roccasBigFivePersonality2002, rothmannBigFivePersonality2003}.
Those who score low on Neuroticism tend to handle stressful situations well and to be calm, poised, and emotionally stable  \cite{roccasBigFivePersonality2002, rothmannBigFivePersonality2003}.

\item \textbf{Openness (O)}
Those who score high on openness tend to be open-minded, imaginative, intellectually curious, and independent of judgment from others~\cite{roccasBigFivePersonality2002, rothmannBigFivePersonality2003}.
Those who score low on openness tend to be down-to-earth, shy away from novelty, and seek conformity \cite{roccasBigFivePersonality2002, rothmannBigFivePersonality2003}.

\end{itemize}

The traits can be interpreted in isolation or in combination with each other and aggregated into \textit{personality types} \cite{sava2011personality}.

\subsection{Personality Studies of Software Developers and Users}
Beyond psychology, personality types have been in the focus of software engineering research to investigate effects on various phenomena such as pair programming, team effectiveness, individual performance, software task allocation, behavior preferences, education, and project management effectiveness~\cite{cruzFortyYearsResearch2015}.
However, only a few studies have investigated the personality of software users in contrast to software producers (developers, managers).
Stachl et al.~\cite{stachlPredictingPersonalityPatterns2020} correlated personality types with mobile app usage.
Their results show that the personality type can be predicted by looking at the frequency, duration, timing of app usage, and feature usage such as average text length in messages.
This suggests that the identification of  personality types does not require users to participate in personality tests.
When submitting feedback, a more realistic scenario is to analyze usage data than asking feedback providers to take a personality test.

Various studies in human-computer interaction research examined the effects of personality on user interface preferences.
Alves et al.~used the FFM scale to show a difference in the preferences of GUI styles between different personality types \cite{alvesIncorporatingPersonalityUser2020}.
The authors found that the personality traits have an effect on preferences of font size, information presentation density, and color themes.
A study by Sarsam and Al-Samarrie ~\cite{sarsamFirstLookEffectiveness2018} shows that users are more satisfied with the usage of a learning app if it is designed in accordance with the preferences associated with their strong personality traits.

We are unaware of studies that have examined the impact of personality on user feedback. To the best of our knowledge, the closest work to our is by Tizard et al.~\cite{tizardVoiceUsersDemographic2020}, who studied the demographic properties of feedback providers.
The authors surveyed 1040 software users about their feedback habits, software usage, and demographic information such as gender, age, ethnicity, education, and employment status.
Their results show that usage duration and gender have a significant impact on the amount and frequency of feedback, as well as the positivity of feedback. However, they did not investigate the impact of personality on the provided feedback.

\section{Methodology}\label{sec:method}

\subsection{Study Design}
Our study consisted of two parts: one for collecting user feedback and one for identifying the personality traits of the feedback providers. 
The questionnaires for each part were available to students independently throughout the last two weeks of the semester.
Participants were able to complete either or both of them in any order. Each part lasted for about 15 minutes.

In the first part, participants were asked to review (in text form) and rate (from 1-dissatisfied to 5-satisfied) several Moodle features. 
Moodle is an open-source learning platform designed to cover all aspects of teaching for educators, administrators, and students.
The software is usually self-hosted by institutions and has been the main teaching platform at our department since 2019.
We collected feedback on five Moodle features that were essential for our course: 

\begin{itemize}
\item \textbf{Material overview} is the main page of the course where all the items such as course material, assignments, and additional information are listed and grouped by week and topic.

\item \textbf{Forum} is a StackOverflow-like, simplified, Q\&A thread-feature, that allows students to ask questions and get answers from teachers or peers, vote on answers, and mark questions as resolved.

\item \textbf{Notification} feature notifies about incoming chat messages, announcements, and course calendar deadlines: either on the Moodle web interface or via email.

\item \textbf{Code runner} is an IDE-like environment that enables teachers to create small programming assignments and the students to write and check code.
The text editor in the \textit{code runner} only provides the students with a minimalistic text field without any of the usual IDE features such as auto compilation, code completion, or auto formatting.
Before submitting the code, students can check the correctness of their program by executing the reference unit test.
The code runner records each submission attempt so the number, duration and score is available for evaluation.
Each week the course exercise consisted of 5-10 sub-tasks which must be completed within a week of publication and can be re-submitted indefinitely before the deadline.

\item \textbf{Quiz} supports the creation of different types of questions (such as multiple/single choice, drag and drop, and gap filling) that are solved by students and submitted in the same manner as the programming tasks.
Both code runner and quiz are used by students for homework assignments that are graded automatically.

\end{itemize}
We focused on those features as they were regularly used during the course: including solving weekly tasks, accessing learning material, and communicating with peers and teachers.

We instructed the participants to explicitly state if they don't have any written feedback on a particular feature to prevent them from just skipping questions without reading.
The written feedback questions included one part regarding problems with the respective feature and one regarding wishes and requests. 

The second part included the questions on the Big Five personality traits using the NEO-FFI-30 questionnaire~\cite{koernerPersoenlichkeitsdiagnostikMitNEOFuenfFaktorenInventar2008}.
We chose the 30 question version of the NEO-FFI to reduce the overall length of the study and increase the willingness to participate.
Both parts included the same demographic questions at the end.
So participants could correct or complement incorrect or incomplete answers from the other part.
All questions except the personality test were optional to make the study shorter and potentially attract more participants.

\subsection{Data Collection and Preparation}
The study participants were university students from a large introductory software engineering course at our department.
The course is mandatory for first semester informatics students. It is also attended by students of other majors.
We advertised the study during the main course session as well as in the main teaching platform Moodle.
As incentive, we raffled online shopping vouchers, which participants could receive by completing both parts of the study.
The participation was completely voluntary and did not have any impact on the course performance or grading. 
Before the study, we collected the agreement of a local ethics committee at the university.   
Out of 124 total participants who enrolled in the study, 56 have completed both parts with all mandatory answers to calculate their personality traits.

Before answering the research questions, we labeled the feedback into different categories and score the personality tests as described in the following.

\subsubsection{Feedback Labeling}
Before the labeling, the first author read every piece of feedback to better understand the underlying data. We observed that most comments have multiple types of information.

To label the information types in the feedback, we used a labeling scheme similar to Pagano and Maalej~\cite{paganoUserFeedbackAppstore2013}.
One of the authors manually labeled the feedback with 10 non-exclusive, basic information types \textit{praise}, \textit{criticism}, \textit{shortcoming}, \textit{improvement\_request},  \textit{bug\_report},  \textit{feature\_request}, \textit{content\_request}, \textit{other\_app}, \textit{noise}, and \textit{rationale}.

\textit{Praise}, \textit{criticism} are generic categories for feedback that is positive or negative but non-informative to developers.
\textit{Shortcoming} is a description of something that is wrong with the software, but without stating a  what can be improved.
\textit{Improvement\_request} and  \textit{feature\_request} are similar to \textit{shortcoming}, but with more specific information on what can be improved or added to the software.
\textit{Bug\_report} is the description of a bug or defect in the software that is obviously not intentional.

We added \textit{content\_request} to differentiate requests that are not concerning the software features but rather the course content.
\textit{other\_app} denotes that another software was mentioned or compared to Moodle in the feedback (e.g. by more experienced users).
\textit{Rationale} is a category that denotes the presence of justification or elaboration of the feedback.
For example, if a feedback text contains a \textit{feature\_request} and additionally the participant describes why this feature is needed in their context, we added the label.

\subsubsection{Personality Test Scoring}

We have scored the personality test according to the NEO-PI-R/NEO-FFI manual by Ostendorfer and Angleitner~\cite{ostendorf2004neo}.
Each participant received a score in the interval [0, 50] for each personality trait.
When taking a closer look at the answer patterns, we found that 3 participants answered \textit{disagree} or \textit{partially disagree} more frequently than expected.

We also observed that the answers of 10 participants exhibit concerning pattern.
8 participants have long streaks of answering questions with the same value, including two participants who also have agreed or disagreed with the questions more frequently than expected.
Additionally, two participants have disagreed with the questions more frequently than expected.
The NEO-PI-R/NEO-FFI manual advises treating these cases with caution, but has no strong advice to discard them entirely.
We conducted our subsequent correlation analyses with and without these 10 cases and did not find any notable difference.


\section{Results}\label{sec:results}




We first analyzed the demographic data of our participants.
We collected data about \textit{age}, \textit{gender}, \textit{semester}, and \textit{study program}.
The average participant in our sample is 20 years old and in their first study year.
However, a large portion of participants did not specify their age or study duration.
21 participants are female, 18 participants are male, and 17 did not specify any gender.
43 participants are studying computer science  and 13 have other majors.

\begin{table}[htbp]
\caption{Statistical parameters of personality traits in our sample compared to benchmark by Körner et al.~\cite{kornerDeutscheNormierungNEOFunfFaktorenInventars2008}}
\begin{center}
\begin{tabular}{|l|c|c|c|c|c|}
\hline
 {Traits } &  A &  C &  E &  N &  O \\
\hline
MEAN from~\cite{kornerDeutscheNormierungNEOFunfFaktorenInventars2008} &  30.49 &  32.53 &  26.46 &  19.47 &  24.53 \\
MEAN       &  34.43 &  35.11 &  25.86 &  22.29 &  31.74 \\
SD from~\cite{kornerDeutscheNormierungNEOFunfFaktorenInventars2008}   &   5.68 &   6.59 &   5.92 &   7.41 &   5.54 \\
SD         &   7.35 &   7.14 &   7.49 &  12.04 &   8.00 \\
\hline
\end{tabular}
\label{tab:pers_mean_sd}
\end{center}
\end{table}

\begin{figure}[b]
\centerline{
    \includegraphics[width=\columnwidth]{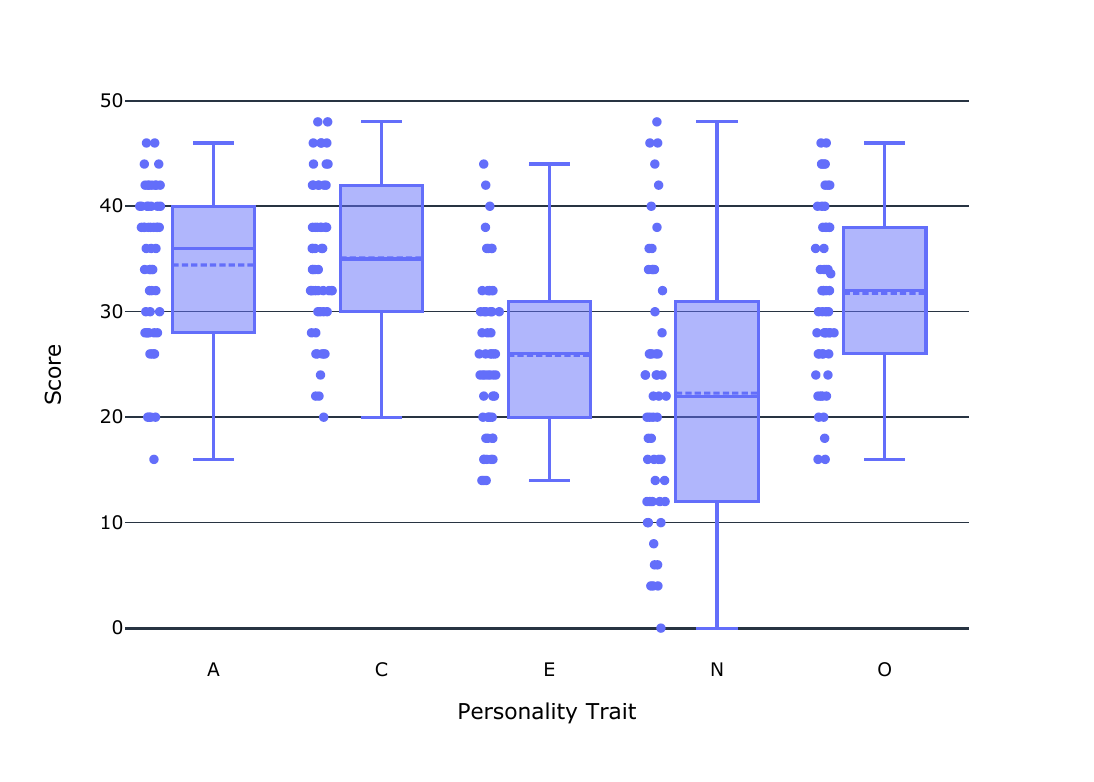}
}
\caption{Distribution of personality scores (N=56).}
\label{fig:personality}
\end{figure}

To simplify our analysis, we limit ourselves to the evaluation of personality traits and not personality types.
Figure~\ref{fig:personality} shows that \textit{agreeableness}, \textit{conscientiousness}, and \textit{openness} are overall higher than \textit{extraversion} and \textit{neuroticism} in our sample.
The means and standard deviations of personality trait scores, shown in Table~\ref{tab:pers_mean_sd}, are close to the original publication by Körner et al.~\cite{kornerDeutscheNormierungNEOFunfFaktorenInventars2008}.
Therefore, we assume that our sample is close to being representative.
Notably, the mean of \textit{openness} and standard deviation of \textit{neuroticism} have the largest differences to the statistics of Körner et al.
One possible explanation is that people willing to participate in such studies might generally expose a high \textit{openness} score.

\begin{figure}[htbp]
\centerline{
    \includegraphics[width=\columnwidth]{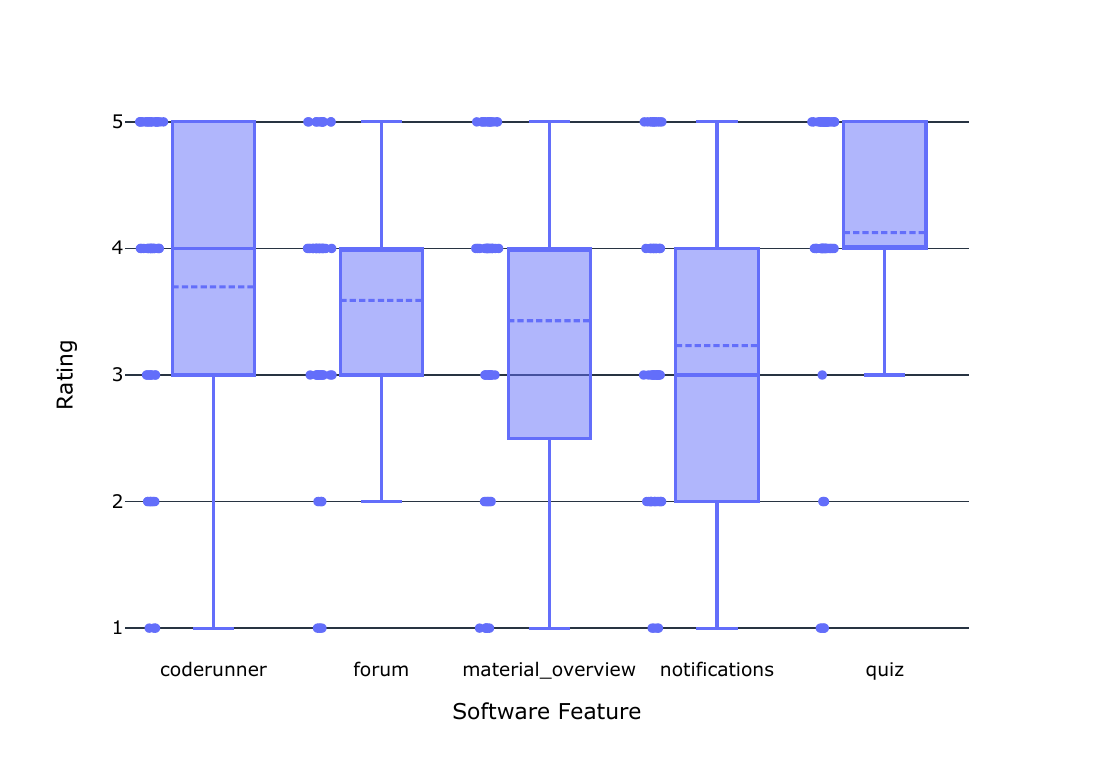}
}
\caption{Distribution of feature ratings (N=56).}
\label{fig:rating}
\end{figure}

Figure~\ref{fig:rating} shows that, on average, the features were rated high, with a median of 4 (solid lines) and a means ranging between 3.23 and 4.13 (dotted lines). 
This corresponds to overall average rating on app store \cite{paganoUserFeedbackAppstore2013}. 
The \textit{quiz} feature was rated highest with a mean of 3.66, the \textit{notification} feature was rated lowest with a mean of 3.23.

Table~\ref{tab:pers_rate_corr_m} shows the results of the correlation analysis conducted to answer RQ1. 
We found one significant correlation between a personality trait and a Moodle feature: namely between \textit{neuroticism} and \textit{material\_overview}(Table~\ref{tab:pers_rate_corr_m}) with a p-value of 0.05.
This means that participants who are high on \textit{neuroticism} (and thus with tendency to experience sadness, embarrassment, anxiety, depression and anger) slightly dislike the \textit{material\_overview}.

\begin{table}[htbp]
\caption{Pearson's coefficient of correlations  between personality trait score and numeric feature ratings (N=56). Bold values mean a significant correlation (p$\leq$0.05)}
\begin{center}
\begin{tabular}{|l|c|c|c|c|c|}
\hline
{Moodle feature} &  A &  C &  E &  N &  O \\
\hline
Material overview &          -0.03 &               0.07 &          0.05 &        \textbf{-0.26} &      0.14 \\
Forum             &          -0.02 &               0.07 &         -0.10 &        -0.16 &      0.01 \\
Notifications     &          -0.08 &               0.15 &          0.09 &        -0.15 &      0.15 \\
Code runner        &          -0.05 &               0.05 &         -0.08 &        -0.16 &     -0.04 \\
Quiz              &          -0.14 &               0.01 &         -0.14 &        -0.07 &     -0.01 \\
\hline
\end{tabular}
\label{tab:pers_rate_corr_m}
\end{center}
\end{table}

\begin{figure}[htbp]
\centerline{
    \includegraphics[width=\columnwidth]{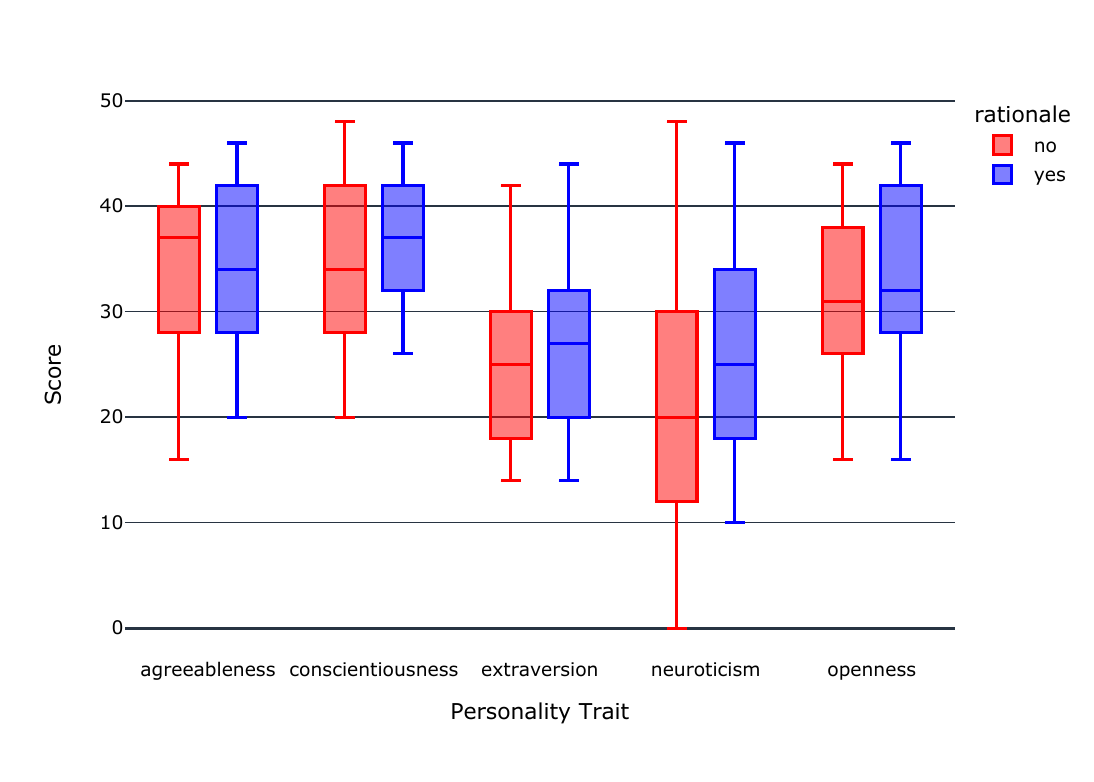}
}
\caption{Rationale provided in user feedback by personality trait (N=56).}
\label{fig:pers_rationale}
\end{figure}

For RQ2, we assessed the elaborateness of written feedback. We particularly looked at the text length, number of mentioned features (Table~\ref{tab:pers_len_corr}), and whether some rationale was provided in the feedback (Figure~\ref{fig:pers_rationale}). 
Overall, the median word length of the feedback is 34 words. The longest written feedback is 220 words long.
We found a positive but not significant correlation between  \textit{neuroticism} and the length of written feedback.
However, \textit{neuroticism} has a significant positive moderate correlation with the number of mentioned features (Table \ref{tab:pers_len_corr}).
This means that in our sample, neurotic individuals tend to write longer feedback, that mentions more features.

\begin{table}[htbp]
\caption{Pearson's correlation between personality traits, feedback length, and number of mentioned features in the feedback (N=56)}
\begin{center}
\begin{tabular}{|l|c|c|c|c|c|}
\hline
{} &  A &  C &  E &  N &  O \\
\hline
{} & \multicolumn{5}{c|}{Feedback length (words)} \\
\hline
correlation &           0.00 &               0.02 &          0.02 &         0.25 &     -0.01 \\
p-value &          0.979 &              0.876 &         0.868 &        0.072 &     0.943 \\
\hline
{} & \multicolumn{5}{c|}{Number of mentioned features} \\
\hline
correlation &          -0.15 &               0.08 &         -0.20 &         \textbf{0.31} &     -0.16 \\
p-value &          0.266 &              0.571 &         0.134 &        0.020 &     0.245 \\
\hline
\end{tabular}
\label{tab:pers_len_corr}
\end{center}
\end{table}

Additionally, we focus on rationale as  RQ2 since it often contains important details for requirements and is an indicator for a deeper involvement by the feedback provider~\cite{kurtanovicMiningUserRationale2017}. 
39 participants provided rationale in their feedback.
Those participants scored higher on personality traits \textit{agreeableness}, \textit{conscientiousness}, \textit{extraversion}, and \textit{openness} and lower on \textit{neuroticism} (Figure \ref{fig:pers_rationale}).

The study data and code is publicly available with an open source license.\footnote{10.5281/zenodo.8173418}

\section{Threats to Validity} \label{sec:threats}

\subsection{External Validity} 
The rather small sample size in our study limits the generalizability of our results and may be the cause of statistical insignificance.
We chose 0.05 as threshold for the $p$-value due to our small sample size, which can impact negatively the validity of our results.

Our participant sample is limited to university students from a beginners programming course and does not represent the general population or the population. Even though our mean and standard deviation of the personality trait scores is similar to the more representative sample, tested by Körner et al.~\cite{kornerDeutscheNormierungNEOFunfFaktorenInventars2008}.

The participants were asked to provide feedback on certain features of the used software. The questionnaire design nudged the participants to differentiate between feedback that voices wishes and feedback that voices problems.
Even though feedback is sometimes explicitly collected in requirements engineering, a big part of it is provided by the users on their own behalf.
Therefore, our results might be limited to a specific type of feedback communication.

A survey-based personality study is prone to over- or underestimation of certain variables.
The willingness to participate in such studies often correlates with high scores on \textit{agreeableness} and \textit{conscientiousness}~\cite{kornerDeutscheNormierungNEOFunfFaktorenInventars2008}.
Our self-selected sample is potentially affected by this.

\subsection{Internal Validity} 
We do not have a control group to test how the feedback behavior might be influenced by other factors, such as experience with user feedback or requirements in general.
We tested for the skill level by looking at the progress of Moodle task completion (duration and number of attempts).
We have no reliable way to detect pauses in the participants' work and can not reliably determine the real duration. 
However, generally, this should not impact the actual tasks in our study (give feedback and answer personality test).

\subsection{Construct Validity} 
We have implemented our personality test questionnaire as instructed by the publication of Körner et al.~\cite{kornerDeutscheNormierungNEOFunfFaktorenInventars2008}. But we did not consult any psychologists prior to the admission of the test.
We performed the scoring and evaluation of the NEO-FFI-30 questionnaire following the instructions from the original publication by Ostendorf and Angleitner~\cite{ostendorf2004neo}.
However, we had no supervision by or consultation with psychologists and can not do an in-depth interpretation of the results.
The questionnaire Körner et al.~\cite{kornerDeutscheNormierungNEOFunfFaktorenInventars2008} is a  reduced version of the original 240 or 60 questions inventory, and therefore possibly less reliable than the original one.


\section{Discussion} \label{sec:discussion}

\subsection{RQ1: Do feature ratings correlate with personality trait
scores?}

Our results show a weak negative correlation between the personality trait \textit{neuroticism} and the  rating of specific Moodle feature \textit{material\_overview}.
This is in line with the study by Alves et al.~who conclude that individuals high on \textit{neuroticism} prefer low information density GUIs.
The Moodle GUI, however, is rather overloaded with different widgets, that are not always necessary for pursuing the course.

We expected to see more correlations between other personality traits as well.
Especially, we expected a high \textit{neuroticism} score to have a much higher negative impact on \textit{code runner} and \textit{quiz} rating, since they were used in a rather stressful course environment.
On the other hand, we expected \textit{forum} and \textit{notification} to be rated higher by individuals who are high on \textit{agreeableness}.
Overall, we did not find a strong universal relationship between  personality traits and rating feedback. But the observed trends should be checked and explored further.

\subsection{RQ2: Do personality trait characteristics affect elaborate-
ness of user feedback?}

The personality trait \textit{neuroticism} has a non-significant, weak positive correlation with the written feedback length, indicating that users who score higher on the \textit{neuroticism} personality trait tend to write longer feedback.
At the same time, the number of mentioned features in their feedback is higher too, which indicates a higher elaborateness and overall engagement.
However, these types of users seem to provide rationale less often than individuals who are high on other FFM scales.
The ones who provide rationale in their feedback tend to have a higher score on the scales \textit{agreeableness}, \textit{conscientiousness} and \textit{extraversion} and particularly \textit{openness}.
This makes sense, especially for high \textit{agreeableness} and \textit{conscientiousness} scores, since they stand for cooperativeness and dependability.
We find some limited evidence that neuroticism may have an impact on the elaborateness of feedback.

\section{Conclusion}\label{sec:conclusion}

Our results show some sporadic effects of personality traits on user feedback.
However, we can not positively answer RQ1 with our result and only have an inconclusive answer for RQ2.
We expected the correlations to be more pronounced for certain personality trait.
We expected participants who score high on \textit{neuroticism} to allocate lower score to Moodle features, as they have used them in a stressful environment.
We expected participants who score high on \textit{conscientiousness}, \textit{agreeableness} to provide longer, more elaborate feedback due to the urge of helping others, inherent to these personality traits.

We assume that the small sample and the specific context of university students could have an impact on the results and would like to repeat the study with a larger, more diverse sample and feedback on a wider range of software.
Regarding the questionnaire, we believe that future work should include mandatory questions about the participants' demographics and previous feedback behavior, to control for those factors when evaluating the impact of personality.
Our results suggest digging deeper into the impact of neuroticism on user feedback, since it is the only personality trait that shows positive results.
Our study focuses on one type of feedback, that is requested from the users (pull feedback).
To get a more complete understanding of the underlying phenomena, it is necessary to analyze feedback that was provided by users on their own behalf.
We believe that analyzing user feedback from public sources and conducting the personality test afterward independently could improve the external validity of the results.

\bibliographystyle{IEEEtran}
\bibliography{references}

\end{document}